\documentclass[twocolumn, twocolappendix]{aastex631}
\usepackage{amsmath}	\usepackage{amssymb}	\usepackage{scrextend}
\usepackage{lipsum,xcolor}
\usepackage{graphicx,midfloat,capt-of}
\DeclareUnicodeCharacter{2212}{-}

\shorttitle{Unveiling Low-Frequency Eclipses in Spider MSPs using wideband GMRT Observations}
\shortauthors{Kumari et al.}

\begin{document}
\title{Unveiling Low-Frequency Eclipses in Spider MSPs using wideband GMRT Observations}
\correspondingauthor{Sangita Kumari}
\email{skumari@ncra.tifr.res.in}

\author[0000-0002-3764-9204]{Sangita Kumari}
\affiliation{National Centre for Radio Astrophysics, Tata Institute of Fundamental Research, S. P. Pune University Campus, Pune 411007, India}
\author[0000-0002-6287-6900]{Bhaswati Bhattacharyya}
\affiliation{National Centre for Radio Astrophysics, Tata Institute of Fundamental Research, S. P. Pune University Campus, Pune 411007, India}
\author[0000-0001-8801-9635]{Devojyoti Kansabanik}
\affiliation{Cooperative Programs for the Advancement of Earth System Science, University Corporation for Atmospheric Research, Boulder, CO, 80301 USA}
\affiliation{NASA Jack Eddy fellow at the Johns Hopkins University Applied Physics Laboratory, 11001 Johns Hopkins Road, Laurel, MD, 20723 USA}
\author[0009-0001-9428-6235]{Rahul Sharan}
\affiliation{National Centre for Radio Astrophysics, Tata Institute of Fundamental Research, S. P. Pune University Campus, Pune 411007, India}
\author[0009-0002-3211-4865]{Ankita Ghosh}
\affiliation{National Centre for Radio Astrophysics, Tata Institute of Fundamental Research, S. P. Pune University Campus, Pune 411007, India}
\author[0000-0002-2892-8025]{Jayanta Roy}
\affiliation{National Centre for Radio Astrophysics, Tata Institute of Fundamental Research, S. P. Pune University Campus, Pune 411007, India}

\begin{abstract}
Eclipses of radio emission have been reported for $\sim$ 58 spider millisecond pulsars (MSP), of which only around 19\% have been extensively studied. Such studies at low frequencies are crucial for probing the properties of the eclipse medium. This study investigates eclipses in 10 MSPs in compact orbit using wide-bandwidth observations with the Giant Metrewave Radio Telescope. We report the first evidence of eclipsing for PSR J2234$+$0944 and J2214$+$3000 in one epoch, while no evidence of eclipsing was observed in the subsequent two epochs, indicating temporal evolution of the eclipse cutoff frequency in these systems. Constraints on the eclipse cutoff frequency were obtained for PSR J1555--2908, J1810+1744, and J2051--0827. Moreover, for the first time, we detected an eclipse at a non-standard orbital phase ($\sim$ 0.5) for PSR J1810+1744, with a duration longer than the eclipse observed at superior conjunction. No eclipses were detected for PSR J0751+1807, J1738+0333, an
d J1807--2459A at 300--500 MHz and 550--750 MHz. We calculated the mass loss rate of the companion for PSR J1555--2908 and PSR J1810+1744, and found that these rates are insufficient to ablate the companion stars. We cataloged the $\dot{E}/a^2$, mass function, roche lobe filling factor, and inclination angle for compact MSP binaries with low-mass companions and found that higher spin-down flux does not guarantee eclipses. Our analysis, supported by the  Kolmogorov–Smirnov statistic, confirms that eclipsing black widow binaries generally exhibit a higher mass function compared to non-eclipsing black widow binaries, consistent with previous studies.

\end{abstract}

\keywords{pulsars: general; binaries: eclipsing, pulsars: individual}

\section{Introduction}
\label{sec:intro}
Millisecond pulsars (MSPs) in compact binary systems with orbital periods less than a day ($P_{b} < 1$) are known as spider MSPs. These systems are categorized into two types based on the mass of their companion: redback (RB) MSPs, with companion masses between 0.1 and 0.9 $M_{\odot}$, and black widow (BW) MSPs, with companion masses less than 0.05 $M_{\odot}$ \citep{roberts2012surrounded}.
Due to the compact orbit, the relativistic pulsar wind ablates the companion star, and this ablated material is believed to cause the frequency-dependent eclipses observed in many of these systems, typically occurring around the superior conjunction (i.e., orbital phase $\sim$ 0.25). Below a certain frequency, defined as the eclipse cutoff frequency \citep[$\nu_{c}$,][]{kansabanik2021unraveling}, the pulsed signal disappears, while it remains detectable at higher frequencies.

With the advent of new telescopes, the population of spider MSPs is steadily increasing, yet eclipses are observed in very few of these systems. For example, PSRs J0023+0924, J2214+3000, J2234+0944 \citep{bak2020timing}, and PSR J0610--2100 \citep{desvignes2016high,J0610-2100_non_Eclipse} are spider MSP systems where no signature of radio eclipses has been seen. 
The lack of eclipses could be due to the lack of low-frequency observations, as eclipses are more prominent at low frequencies. However, for spider MSPs where low-frequency observations exist, the absence of eclipses could be due to the following reasons.
First, the companion star may not be filling its Roche lobe, making it difficult to expel material. For instance, the non-detection of eclipses in PSR J0023+0923 could be attributed to its low roche lobe filling factor (RLFF) of 0.3 \citep{J0023+0923_rochelobe}. Similarly, for PSR J0610$-$2100, the eclipse is likely undetected because the companion is not filling its Roche lobe \citep{J0610-2100_non_Eclipse}.
Second, a low inclination angle of the orbit could also explain the absence of eclipses. For example, PSR J0636+5128 has a large RLFF of 0.75, yet it does not exhibit eclipses, possibly due to its low inclination angle of 24 degrees \citep{J0636+5128}.
Additionally, the absence of eclipses could be due to a low $\dot{E}/a^{2}$ (where $\dot{E}$ is the spin-down energy of the pulsar and $a$ is the distance between the pulsar and the companion). This ratio provides a measure of how effective the isotropic pulsar wind would be at the position of the companion, which in turn could serve as a metric for effective ablation. However, no correlation has been previously studied between the presence of eclipses and the corresponding $\dot{E}/a^{2}$ value, and a larger sample of eclipsing systems is needed to investigate this further. Finally, the lack of eclipses could also depend on the nature of the companion, as many compact systems with white dwarf companions do not exhibit eclipses \citep{J1641+3627D_whitedwarf,J1719-1438_msp}.

Discovering and investigating the eclipse characteristics of more spider systems is of prime importance. Spider MSPs are thought to be the progenitors of isolated MSPs. To date, no spider MSP system has been discovered where it is possible to completely ablate the companion within any reasonable timescale \citep{polzin2020study,stappers1996probing}.
Hence, finding eclipses using low-frequency observations in more spider systems would allow us to calculate the mass loss rate, contributing to the search for such a pulsar. With the advent of wide-bandwidth radio telescopes, precise estimation of eclipse cutoff frequencies has become possible, as demonstrated by \cite{kansabanik2021unraveling} for PSR J1544+4937 using the upgraded Giant Metrewave Radio Telescope \citep[uGMRT][]{gupta2017upgraded}. Determining the cutoff frequencies for other spider populations would enable the study of the probable eclipse mechanisms.
\cite{Thompson1992} suggested various mechanisms to explain the low-frequency eclipses in these systems. For example, cyclotron-synchrotron absorption is believed to be the major eclipse mechanism for PSR J1810+1744 \citep{PolzinJ1810}, J1227$−$4853 \citep{Kudale2020}, and J2215+5135 \citep{J2215-5135_eclipse}, while scattering and cyclotron absorption are considered the primary mechanisms for PSR J2051$−$0827 \citep{PolzinJ2051}. Furthermore, stimulated Raman scattering has been suggested as the most plausible eclipse mechanism for PSR B1744$−$24A \citep{Thompson1992}.
Moreover, \cite{kumari2024} reported temporal variations in the eclipse cutoff frequency for PSR J1544+4937. Assuming that synchrotron absorption is the primary mechanism driving the eclipses for this pulsar, the study by \cite{kumari2024} inferred that changes in the eclipse cutoff frequency might result from variations in factors such as electron density within the eclipse medium, the spectral index of the electron energy distribution, the magnetic field strength, and the orientation of the magnetic field relative to our line of sight. Their study also offered the first evidence that the mass loss rate of the companion influences frequency-dependent eclipsing. Conducting a similar study on other spider MSPs would also contribute to a better understanding of frequency-dependent eclipsing in these systems.

The frequency-dependent nature of the eclipses makes them easier to detect at lower frequencies, where they are more pronounced, compared to higher frequencies. Therefore, the uGMRT is the perfect instrument for studying frequency-dependent eclipsing, as it allows for wide-bandwidth observations at low frequencies ($\sim$ 300 MHz). Additionally, the wide bandwidth provided by uGMRT enables precise estimation of the eclipse cutoff frequency.

In this paper, we present the results from low frequency study using the uGMRT for a sample of compact orbit binary MSPs. Section \ref{sec:obs} presents the details of the observations. Section \ref{sec:analysis} describes the data analysis and Section \ref{sec:results} presents the results and interpretations. Section \ref{sec:possible cause of eclipsing} discusses the possible causes of eclipsing. The summary of the paper is detailed in Section \ref{sec:summary}.

\section{Observations}
\label{sec:obs}
We selected a sample of 10 MSPs that are part of compact binary orbits with orbital period ($P_{b}$) less than a day. These MSPs were expected to be bright enough to be detected with signal-to-noise ratio (SNR) $\ge$ 22 in 2 minutes of integration (considering 13 antenna at 400 MHz, with 200 MHz bandwidth), providing better orbital phase sampling required for the investigation of frequency-dependent eclipsing. All the observations presented in this paper were carried with the uGMRT system \citep{gupta2017upgraded}. GMRT is an interferometeric array consisting of 30 identical dishes, each having 45m of diameter. Since, the location of each of the pulsar in our sample is precisely known through timing, the observations in this study were conducted in coherent phased array mode. The data was recorded at the best possible time and frequency resolution at band 2 (125--250 MHz), band 3 (300--500 MHz), band 4 (550--950 MHz) and band 5 (1050--1450 MHz). In some of the observing epoch
s, we have split the antennas into two sub-arrays ($\sim$ 13 antenna each) at different frequency bands to conduct simultaneous observations. The observing frequency bands and the bandwidth chosen for individual observations is detailed in Table \ref{tab:Table1}. 

\section{data analysis}
\label{sec:analysis}
To mitigate short-duration broad-band radio frequency interference (RFI), we employed the GMRT pulsar tool (gptool\footnote{\label{note1}\url{https://github.com/chowdhuryaditya/gptool}}) software. After the RFI mitigation, we corrected for the interstellar dispersion using the incoherent dedispersion technique (for a few of the epochs, we also utilized the coherent de-dispersion technique). We folded the resulting de-dispersed time series with the known radio ephemeris for pulsars, utilizing the $\it{prepfold}$ task of $\it{PRESTO}$ \citep{ransom2002fourier}. The mean pulse profile was cross-correlated with a high signal-to-noise template profile from past observations to obtain the observed times of arrival (TOAs) of the pulses. TOAs were generated using the Python script $\it get\_TOAs.py$ from $\it{PRESTO}$ \citep{ransom2002fourier}. We calculated the timing residuals, which are the differences between the observed and predicted TOAs, using the $\it{tempo2}$ software packa
ge \citep{hobbs2006tempo2}. The excess dispersion measure (DM\textsubscript{excess}) in the eclipse region (orbital phase $\sim$ 0.2--0.3) introduces an extra time delay, which can be determined using the relation \citep{LorimerKramer}: 

\begin{equation}
\label{Excess_DM}
    \mathrm{DM_{excess} (pc ~ cm^{-3}) = 2.4 \times 10^{-10} t_{excess} (\mu s) f^{2} (MHz)}
\end{equation}
where, $\mathrm{t_{excess}}$ is the excess time delay in the eclipse region in $\mu s$ and f is the observing frequency in MHz.
From $\mathrm{DM_{excess}}$ the electron column density in the eclipse medium ($N_{e}$) is computed using :
\begin{equation}
\label{N_e_determination}
\mathrm{N_{e}(cm^{-2}) = 3 \times 10^{18} \times DM (pc ~ cm^{-3})}   
\end{equation}

\subsection{Eclipse parameters estimation}
\label{subsec:eclipse parameter estimation}
Below, we outline the procedure used to determine the eclipse cutoff frequency and eclipse width for the pulsars in our sample \citep{Sharan_et_al_2024}, which was also employed by \cite{kumari2024}. 

In the baseline subtracted data cube (time, frequency, phase bin), intensity of OFF and ON phase bins, both follow Gaussian random distribution with the same standard deviation ($\sigma$) but with different mean. Hence, for determining ON and OFF phase bins, a quantity, H, is calculated using the relation, H = (sum of N number of samples)/($\mathrm{\sigma \times \sqrt{tf^{'}}}$), for each phase bin, where N ($\mathrm{=tf^{'}}$) is the number of samples, t is the number of sub-integrations for a given timestamp, and $\mathrm{f^{'}}$ corresponds to the number of sub-bands in a frequency chunk.

For the cutoff frequency estimation, in the above expression of H, we have used t to be number of sub-integrations in $\sim$ 5 mins, covering the eclipse phase and f corresponding to the number of sub-bands in frequency for $\sim$ 40 MHz chunk.
The phase bins with an H value greater than 3 (for the fixed time chunk of $\sim$ 5 mins around the eclipse phase) were labeled as ON, otherwise, they were labeled as OFF. This provides the ON and OFF phase bins for each 40 MHz frequency chunk, and based on this, we estimated the cutoff frequency.

For the eclipse radius estimation, in the above expression of H, we have used t corresponding to a sub-integration in time ($\sim$ 1.16 mins for PSR J1810+1744 and 2.40 mins for PSR J1555--2908) and f corresponding to the number of sub-bands in frequency for $\sim$ 28 MHz chunk for PSR J1810+1744 and $\sim$ 60 MHz chunk for PSR J1555--2908. The phase bins with an H value greater than 3 were labeled as ON, otherwise, they were labeled as OFF. We calculated the ON phase bins for each sub-integration in time, keeping the frequency chunk fixed. The SNR of the pulse profile was calculated for every sub-integration following \cite{LorimerKramer}, for each frequency chunk. We defined the eclipse width based on the duration during which the SNR is below 4 for each frequency chunk. 
We thereafter converted this eclipse width calculated in minutes into distance assuming a circular orbit for the pulsars for the calculation of the mass loss rate of the companion. 

We also determined the mass loss rate of the companion using the relation, $\mathrm{\Dot{M_{c}} \sim \pi R_{E}^{2} m_{p} n_{e} V_{w}}$ \citep{Thompson1992,PolzinJ1810}, where $\mathrm{R_{E}}$ is the eclipse radius, $\mathrm{m_{p}}$ is the mass of the proton, $\mathrm{n_{e} = N_{e}/(2R_{E})}$, is the electron volume density, and $\mathrm{V_{w} = (U_{E}/(n_{e} m_{p}))^{1/2}}$, is the velocity of the material entrained in the pulsar wind. Here, $\mathrm{U_{E} = \Dot{E}/(4 \pi c a^{2})}$ represents the energy density of the pulsar wind, with $\Dot{E}$ being the spin-down energy and $\mathrm{a}$ being the distance between the pulsar and the companion. The mass loss rate has been calculated under the assumption that material is spherically symmetric around the companion. The corresponding values of the excess electron density, eclipse cutoff frequency, eclipse radius and mass loss rate are reported in Table \ref{tab:Table1}.

\begin{table*}[!htb]
\begin{center}
\caption{Summary of the observations}
\label{tab:Table1}
\vspace{0.3cm}
\label{discovery}
\begin{tabular}{|l|l|l|l|l|l|l|l|l|l|l|}
\hline
Pulsar name     & P$^{a}$ & DM$^{b}$ & $P_{b}^{c}$ &Freq$^{d}$& OP$^{e}$ & Date & $\nu_{c}$ & $N_{e}^{f}$ & $R_{E}^{g}$& $M_{c}^{h}$\\
     & (ms)& ($\mathrm{pc~cm^{-3}}$)& (hr) & &  &  & (MHz) & $\mathrm{(10^{16}cm^{-2})}$  &(cm) & ($\mathrm{M_\odot/year}$)\\
\hline
 
J0751+1807 & 3.47 &30.24& 6.31 & B34 &  0.16-0.31& 03 December 2022 &-   & - & -&-\\

\hline
J1431--4715 & 2.01 &59.35& 10.79 & B34 & 0.20-0.31 & 25 November 2022  & -  &-  & -&-\\
  &   & &  & B3 & 0.22-0.36 & 04 November 2023  & -  & - &- &-\\
\hline 
J1555--2908& 1.78 &75.91&  5.6 & B34  & 0.20-0.38 & 29 June 2020  &  $>$ 740 &  7.3  &- &-\\
 &  &&   & B45  &  0.13-0.55 &  04 July 2020 & $>$ 800 &  11.7 &3.9 $\times 10^{10}$& 8.2 $\times 10^{-13}$\\
\hline  
J1738+0333& 5.85&33.77&  8.51  & B34 &  0.08-0.27& 9 January 2023 & - &  -& -&-\\
\hline
J1807--2459A&3.05&134.00&  1.70   & B32 &  0.47-1.19& 03 June 2020 & - &  & -&-\\
 & & &      & B3  &  0.01-0.73 & 11 June 2020$_{1}$  & - &  - & -&-\\
  & & &      & B3  &  0.80-1.42 & 11 June 2020$_{2}$  & - &  - & -&-\\
  & & &      & B45  & 0.89-1.57 & 27 June 2020  & $<$ 550 &  -  & -&-\\
\hline
J1810+1744  & 1.66 & 39.70 & 3.60 &  B45 &  0.12-0.82 & 26 June 2020$_{1}$&  $>$ 800  & 7.6  &7.5$\times 10^{10}$& 9.0$\times 10^{-13}$\\
   & & & &  B45 & 0.94-1.64& 26 June 2020$_{2}$&  $>$ 800 & 4.3 & 7.5$\times 10^{10}$ & 6.5$\times 10^{-13}$\\
\hline
J2051--0827&4.50&20.73& 2.37& B45 &  0.12-0.75 & 19 June $2020_{1}$&  $<$ 640   & 10.9 &  -&-\\
&&& &  &  0.85-1.82 & 19 June $2020_{2}$& $<$ 640  & 14.4  &- &-\\
 & & &  &   & 0.92-1.48  & 19 June $2020_{3}$& $<$ 640  & 16.9 & - &-\\
\hline
 J2214+3000  &3.11&22.54&  9.99 &  B34& 0.20-0.26 & 11 November 2022& - &  & & \\
    & & &    &  B34& 0.19-0.33 & 30 October 2023& - & - & - &- \\
    & & &    &  B34& 0.18-0.32 & 06 November 2023& -& -  & - & -\\
\hline
J2215+5135   &2.61&69.20&   4.14  & B34 &0.07-0.43 & 11 November 2022 &- &  -&- &-\\
    & & &      &  B3 & 0.03-0.37 & 29 October 2023 &- & - &- &-\\
& & &      &  B3 & 0.19-0.31  & 04 November 2023 & - & - &- &-\\
\hline
J2234+0944   &3.62&17.83&  10.07     &B34 & 0.22-0.28 & 11 November 2022  & - & - & -&-\\
 & & &      & B34 & 0.20-0.34 & 29 October 2023& -  & - &- &-\\
    & & &      & B34 & 0.18-0.32 & 05 November 2023& - & - &- &-\\
\hline

\end{tabular}
\end{center}
$^{a}$: Spin period of the pulsar; $^{b}$: Dispersion measure; $^{c}$: Orbital period of the pulsar; $^{d}$: Observing frequency, B3 - band 3, B34 - simultaneous band 3 (300--500 MHz) and band 4 (550--750 MHz) observations with 200 MHz bandwidth, B32 - simultaneous band 3 (400--500 MHz) and band 2 (200--300 MHz) observations with 100 MHz bandwidth, B45 - simultaneous band 4 (550--950 MHz) and band 5 (1050--1450 MHz) observations with 400 MHz bandwidth; $^{e}$: Orbital phase covered; $^{f}$: The maximum electron column density in the eclipse medium; $^{g}$: Eclipse radius; $^{h}$: Mass loss rate of the companion.

\vspace{1cm}
\end{table*} 

\section{Results and Interpretation}
\label{sec:results}
Based on the analyses reported in Section \ref{sec:obs}, we divide the target MSPs into four categories. First, MSPs for which no eclipse is reported in previous studies, but for which we observe evidence of eclipsing. Second, MSPs where an eclipse is reported in previous studies, and our aim is to determine the eclipse cutoff frequency. Third, MSPs for which no eclipse is reported in previous studies, and we also do not find eclipsing using low-frequency observations with the uGMRT. The last category includes MSPs that are not detected at low frequencies in our sample. We report the results for each category in detail in the following sections.

\subsection{First evidence of eclipsing}
We report the first evidence of eclipsing for PSR J2214+3000 and PSR J2234+0944. Below, we will discuss the results obtained for each pulsar in detail. 

\subsubsection{ PSR J2214+3000}
\begin{figure*}
     \centering
     \begin{minipage}[b]{0.38\textwidth}
         \centering
         \includegraphics[width=\textwidth]{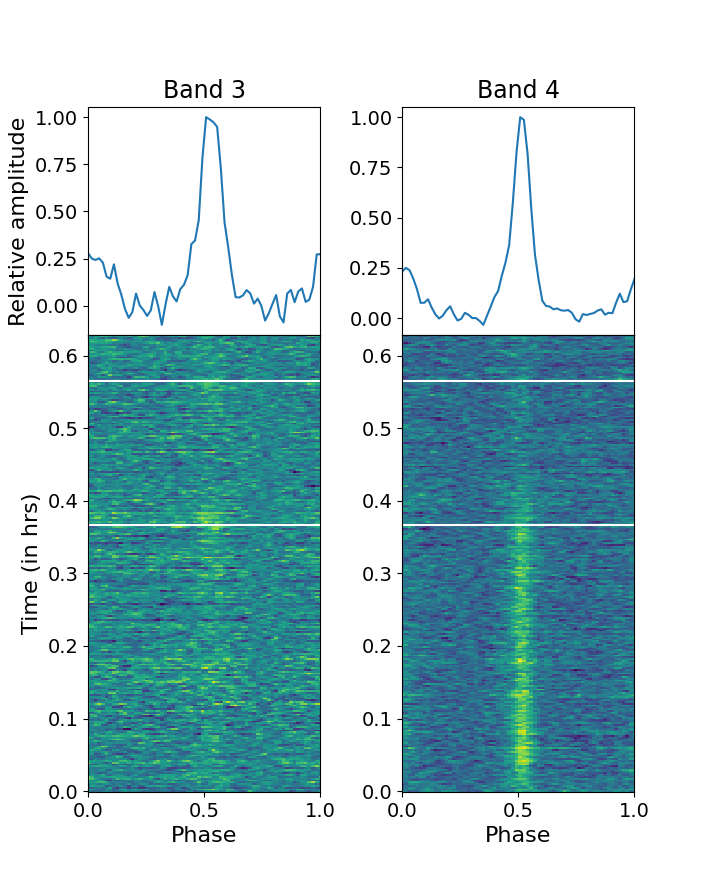}
         \textbf{(a)} Phase versus time plot for PSR J2214+3000 for simultaneous band 3 and band 4 observations on 11 November 2022. The white lines covers the orbital phase from 0.23$-$0.26. The eclipse can be seen in band 3, and flux fading can be observed in band 4.
     \end{minipage}  
     \hspace{0.05\textwidth}
     \begin{minipage}[b]{0.38\textwidth}
         \centering
         \includegraphics[width=\textwidth]{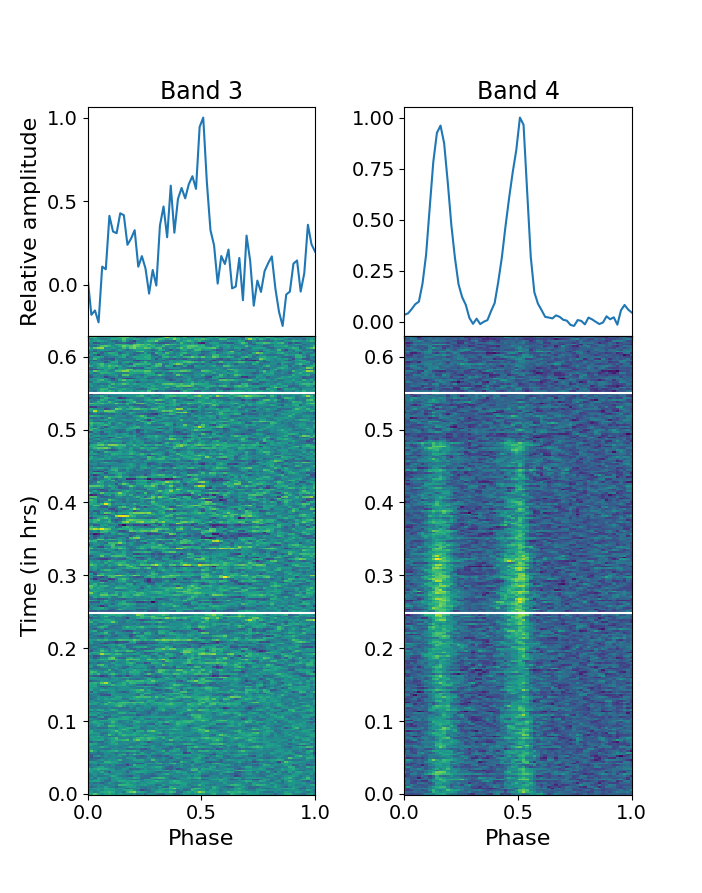}
         \textbf{(b) }Phase versus time plot for PSR J2234+0944 for simultaneous band 3 and band 4 observations on 11 November 2022. The white lines covers the orbital phase from 0.25$-$0.28. Eclipse is observed in band 3; however, no eclipse is noted in band 4.
     \end{minipage}
    \caption{Detection plot showing the first evidence of eclipsing for PSR J2214+3000 and PSR J2234+0944.}
        \label{both_J2214_J2234_plots}
\end{figure*}

We analyzed the simultaneous band 3 and band 4 data near the eclipse phase (orbital phase $\sim$ 0.2$–$0.3), corresponding to three epochs, to investigate the presence of eclipsing in the PSR J2214+3000 system (see Table \ref{tab:Table1}). Figure \ref{both_J2214_J2234_plots}a shows the detection plots for 11 November 2022. 

In the previous study by \cite{bak2020timing} on PSR J2214+3000, no evidence of eclipse was observed using WSRT, Effelsberg, Lovell, Nan\c{c}ay Radio Telescope covering the frequency range 385 MHz to 1.5 GHz. \cite{bak2020timing} also did not observe any variation in DM near the superior conjunction (orbital phase $\sim$ 0.25), which could be a result of coarser orbital phase sampling.

For the first time, we found the signature of frequency-dependent eclipsing in PSR J2214+3000 in the observation on 11 November 2022. A clear signature of eclipsing can be seen in Figure \ref{both_J2214_J2234_plots}a, as the pulsar is not detected in the eclipse phase from 0.23 to 0.26 (shown by the white lines in the Figure) in band 3, whereas some flux fading is observed in band 4 during those orbital phases. However, in other epochs of observation, no signature of eclipsing has been noted. This clearly shows that the eclipse environment for this MSP is dynamically evolving, as seen in other BW MSP systems \citep[for example, as seen for PSR J1544+4937 by][]{kumari2024}.

This pulsar is part of the European Pulsar Timing Array \citep[EPTA,][]{bak2020timing}, and the detection of eclipses in this system may have implications for the EPTA, as the eclipse phase should be neglected while searching for gravitational wave signals.

\subsubsection{PSR J2234+0944}
We analyzed simultaneous band 3 and band 4 observations for PSR J2234+0944 as listed in Table \ref{tab:Table1}. 
The pulsar is detected in band 4 in all three epochs of observations. However, it is not detected in the simultaneous band 3 observations on 11th November 2022. This non-detection in band 3 could also be due to ionospheric scintillation. In the other two epochs, the MSP is detected in band 3 observations as well. The non-detection of PSR J2234+0944 in band 3 on 11th November 2022, alongside its detection in simultaneous band 4 observations, may suggest the presence of frequency dependent eclipsing in this system. The detection of an eclipse in one epoch and the absence of an eclipse in subsequent epochs may suggest time variable frequency-dependent eclipsing for this pulsar as well, as noted for PSR J1544+4937 by \cite{kumari2024}. Figure \ref{both_J2214_J2234_plots}b depicts the detection for 11 November 2022.

In previous studies, no eclipse has been reported for this MSP by \cite{bak2020timing} using the observations spanning from 385 MHz to 1.5 GHz. \cite{bak2020timing} also studied the orbital phase-dependent DM variations for this MSP (Figure 5 of their paper) and did not find any DM increase near the superior conjunction (orbital phase $\sim$ 0.25), which also indicated the absence of eclipsing material. However, they suggested that this could be due to the coarse sampling of DM over the orbital period, making the variation around the superior conjunction not evident. This pulsar is also a part of EPTA \citep{bak2020timing} and the eclipse phase should be neglected while searching for gravitational wave signal.

\subsection{Determination of eclipse cutoff frequency}
Eclipsing is reported for PSR J1555--2908, PSR J1810+1744, and PSR J2051--0827 in previous studies, and we aim to determine the eclipse cutoff frequency for these pulsars. Below, we will discuss the results obtained for each of these MSPs in detail.

\subsubsection{PSR J1810+1744}

We analyzed simultaneous band 4 and band 5 observations data for PSR J1810+1744 as shown in Table \ref{tab:Table1}. Previously eclipse has been reported upto 750 MHz by \cite{PolzinJ1810}. Figure \ref{J1810_orbital_phase_resloved_plots} shows the detection plots for two consecutive eclipses on 26 June 2020. The pulsar is detected in band 4, where a complete eclipse is observed in both orbits. However, the pulsar is not detected in simultaneous band 5 observations, allowing us to set an upper limit on its flux density at 1.4 GHz to 30 $\mu$Jy, corresponding to a 5-sigma detection significance over 1 hour of integration time. Based on this we conclude that the cutoff frequency would be above 800 MHz (the upper edge of band 4) for this pulsar. Figure \ref{J1810_orbital_phase_resloved_plots}b depicts the orbital phase DM and flux density plot for the second eclipse observed. 

\subsubsection{Eclipse at non-standard orbital phase for PSR J1810+1744}

\begin{figure*}
     \centering
     \begin{minipage}[b]{0.76\textwidth}
         \centering
         \includegraphics[width=\textwidth]{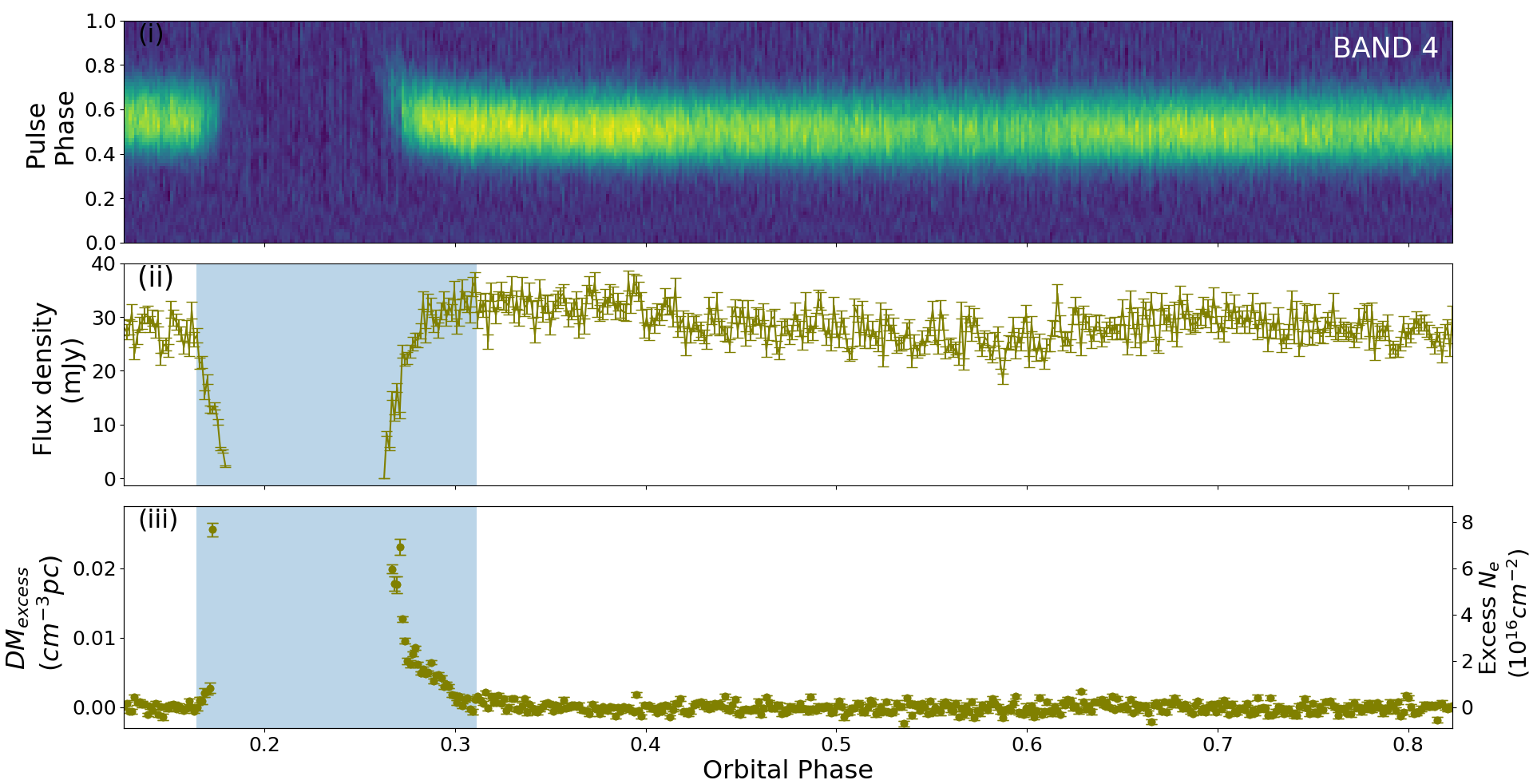}
         \textbf{(a)} First eclipse for J1810+1744 
         
     \end{minipage}
     \hfill
     \begin{minipage}[b]{0.8\textwidth}
         \centering
         \includegraphics[width=\textwidth]{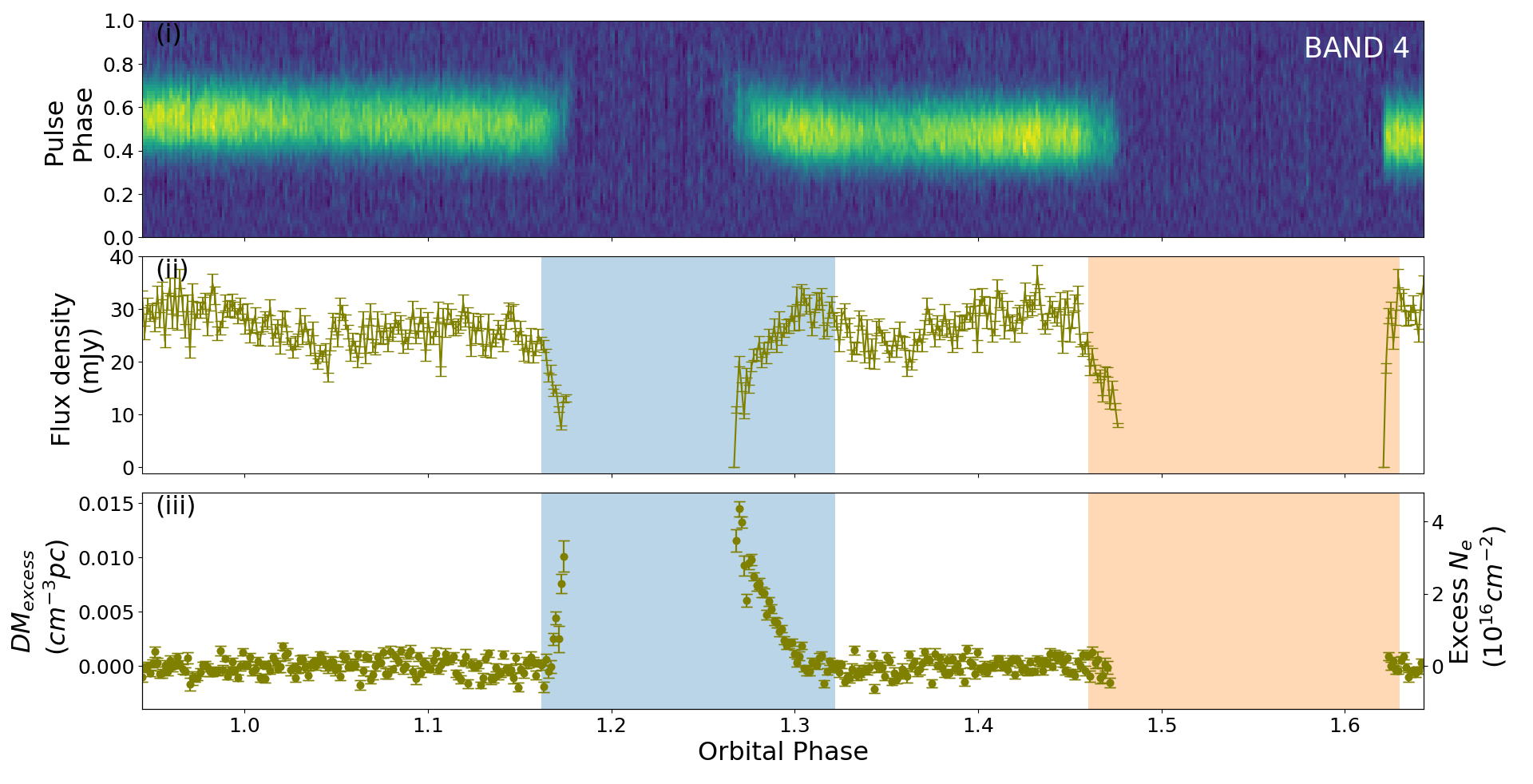}
         \textbf{(b)} Second eclipse for J1810+1744
         
     \end{minipage}
    \caption{ The consecutive eclipses for PSR J1810+1744 on 26 June 2020. Three panels, from top to bottom depict variation of (i) the total intensity at band 4 (ii) flux density of total intensity at band 4(iii)  excess DM ($DM_{excess}$) along the line of sight at band 4. The main eclipse is indicated by the shaded blue region, and the eclipse at orbital phase $\sim$ 0.55 is shown by the orange shaded region.}
        \label{J1810_orbital_phase_resloved_plots}
\end{figure*}

For the second orbit covered, the presence of an eclipse at a non-standard orbital phase (orbital phase $\sim$ 0.5, shown by the orange shaded region in Figure \ref{J1810_orbital_phase_resloved_plots}b) is also seen, along with the main eclipse around the superior conjunction (orbital phase $\sim$0.25). The duration of the eclipse at 0.5 is $\sim$ 29 mins, which is longer than the eclipse duration at superior conjunction. Eclipses at phases other than superior conjunction have been observed for other BW MSPs in the past, but not for a duration longer than the main eclipse. For example, a mini eclipse has been reported for PSR J1717+4308A using the FAST telescope \cite{J1717+4308A_eclipse} and for PSR J1544+4937 by \cite{bhattacharyya2013gmrt}. 
We also note from Figure \ref{J1810_orbital_phase_resloved_plots} that, for the eclipse occurring at an orbital phase around 0.5, no rise in the DM is observed near the eclipse boundary. A similar behavior was also noted for PSR J1717+4308A \cite[Figure 1 of][]{J1717+4308A_eclipse}, where no time delay in pulse arrivals was seen at the mini-eclipse boundary.
An eclipse other than at superior conjunction (orbital phase $\sim$ 0.25) has not been reported for PSR J1810+1744 before, and this study provides the first detection of such an event. The occurrence of an eclipse at a non-standard orbital phase is unusual, as it was not observed in the first orbit but suddenly appeared in the next orbit, with a duration longer than that of the main eclipse. This indicates that the system is highly dynamic, with additional material suddenly appearing and extending over a larger range at the orbital phase of 0.5 compared to the main eclipse phase ($\sim$ 0.25). The cause of this eclipse could be the anisotropic and clumpy distribution of ionized gas in the orbit, which could be probed with regular follow up of the system.

\begin{figure*}
     \centering
     \begin{minipage}[b]{0.8\textwidth}
         \centering
         \includegraphics[width=\textwidth]{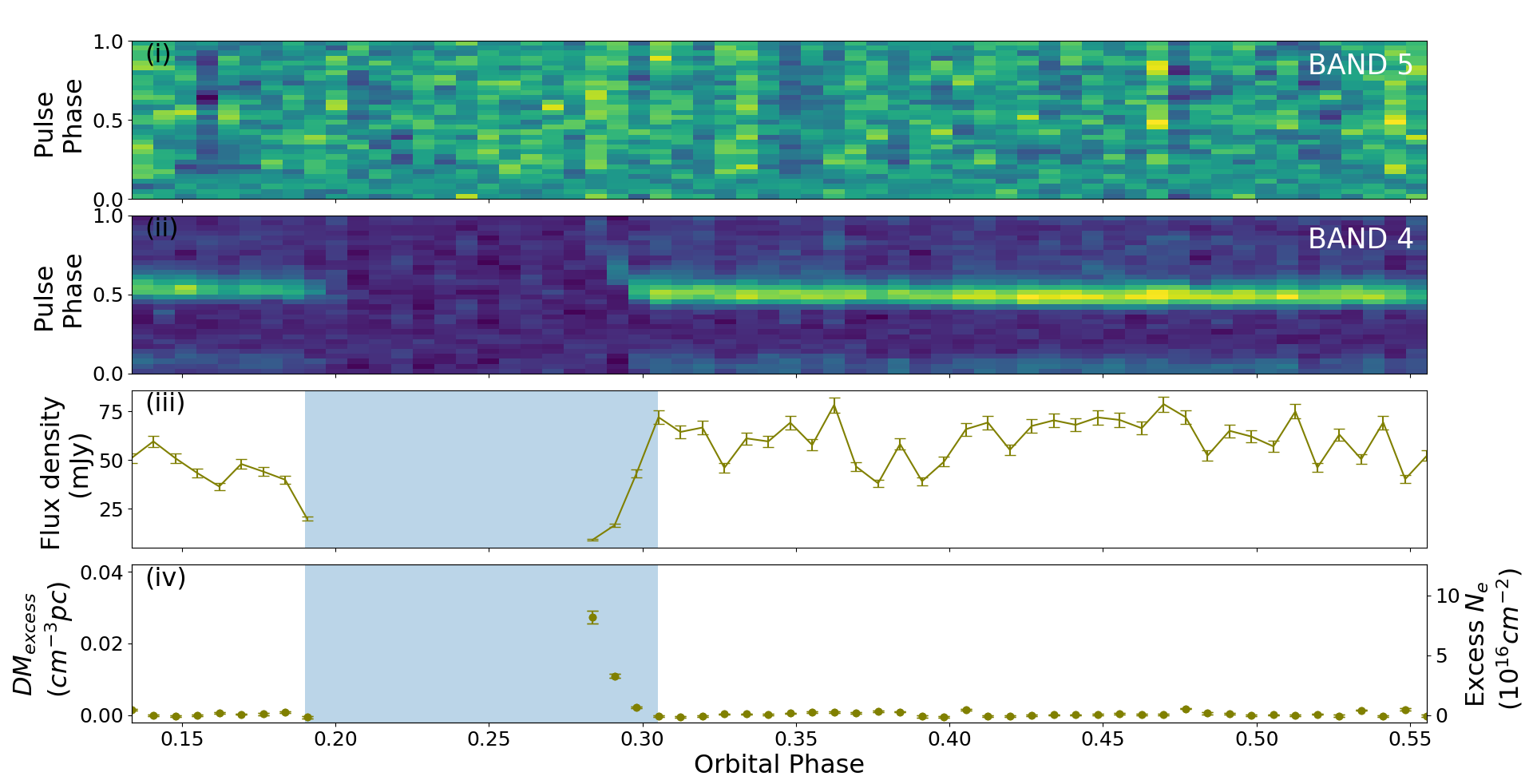}
         \textbf{(a)} 04 July 2019 plots for PSR J1555$-$2908
         
     \end{minipage}
     \hfill
     \begin{minipage}[b]{0.8\textwidth}
         \centering
         \includegraphics[width=\textwidth]{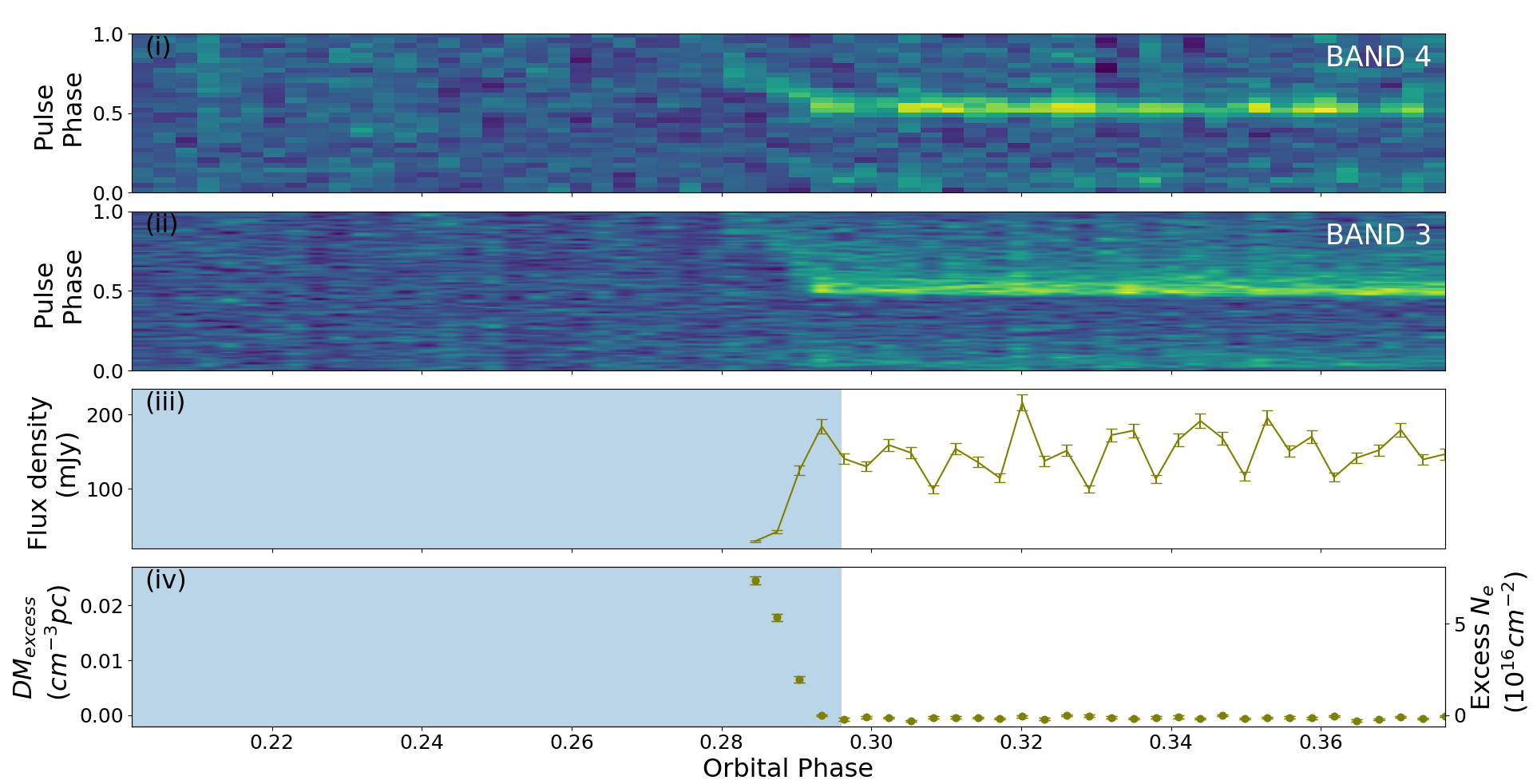}
         \textbf{(b)} 29 June 2019 plots for PSR J1555$-$2908
         
     \end{minipage}
    \caption{Detection plots for PSR J1555$-$2908 for two different epochs. a) Four panels, from top to bottom depict variation for 4 July 2019 of (i) the total intensity at band 5, (ii) the total intensity at band 4, (iii) flux density of total intensity at band 4, (iv) excess DM ($DM_{excess}$) along the line of sight. b) Four panels, from top to bottom depict variation for 29 July 2019 of (i) the total intensity at band 4, (ii) the total intensity at band 3, (iii) flux density of total intensity at band 3, (iv) excess DM ($DM_{excess}$) along the line of sight. The eclipse phase is indicated by the shaded blue region.}
        \label{J1555_orbital_phase_resolved_plots}
\end{figure*}

\begin{figure*}
     \centering
     \begin{minipage}[b]{0.7\textwidth}
         \centering
         \includegraphics[width=\textwidth]{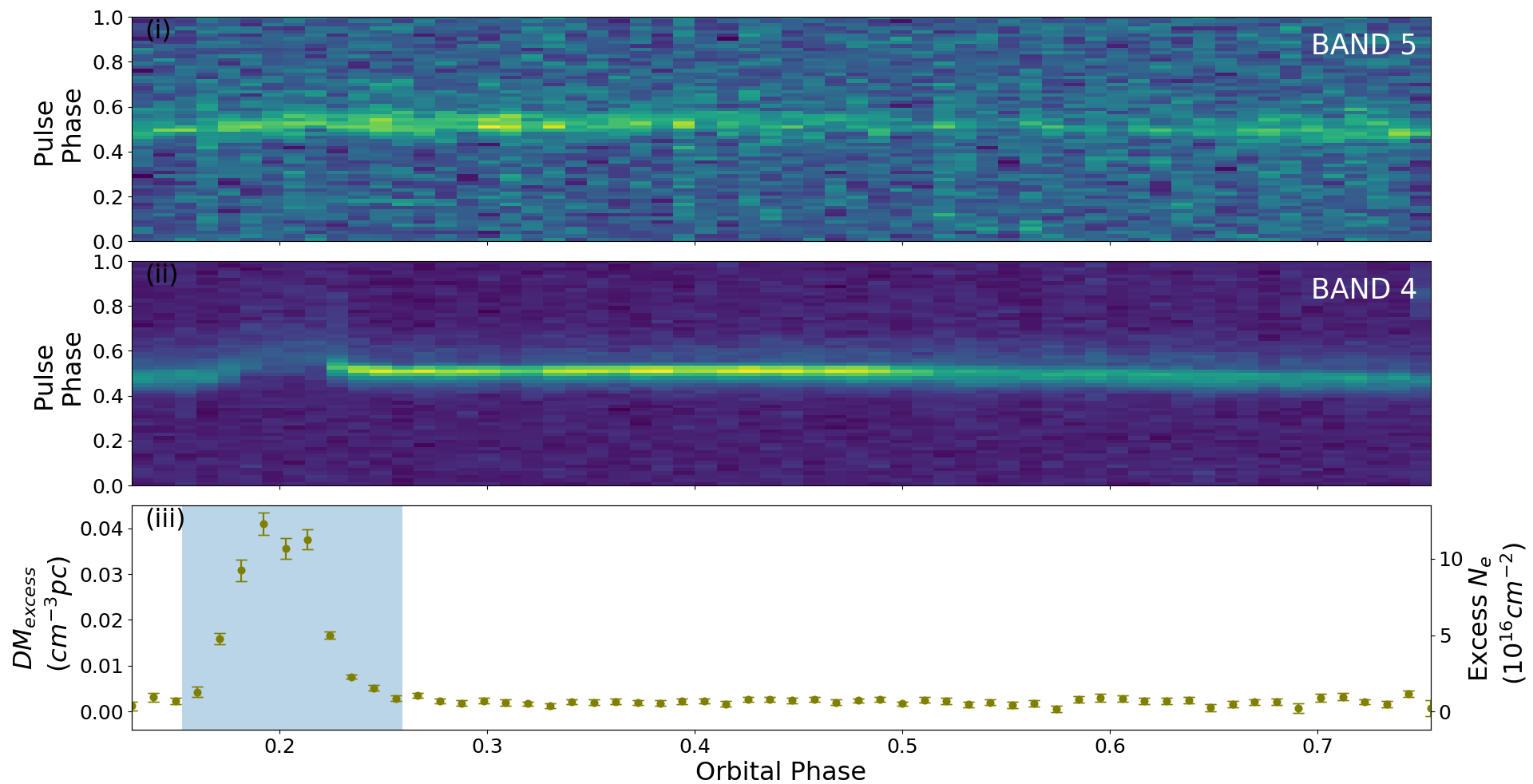}
         \textbf{(a)} First eclipse for PSR J2051$-$0827
         
     \end{minipage}
     \hfill
     \begin{minipage}[b]{0.7\textwidth}
         \centering
         \includegraphics[width=\textwidth]{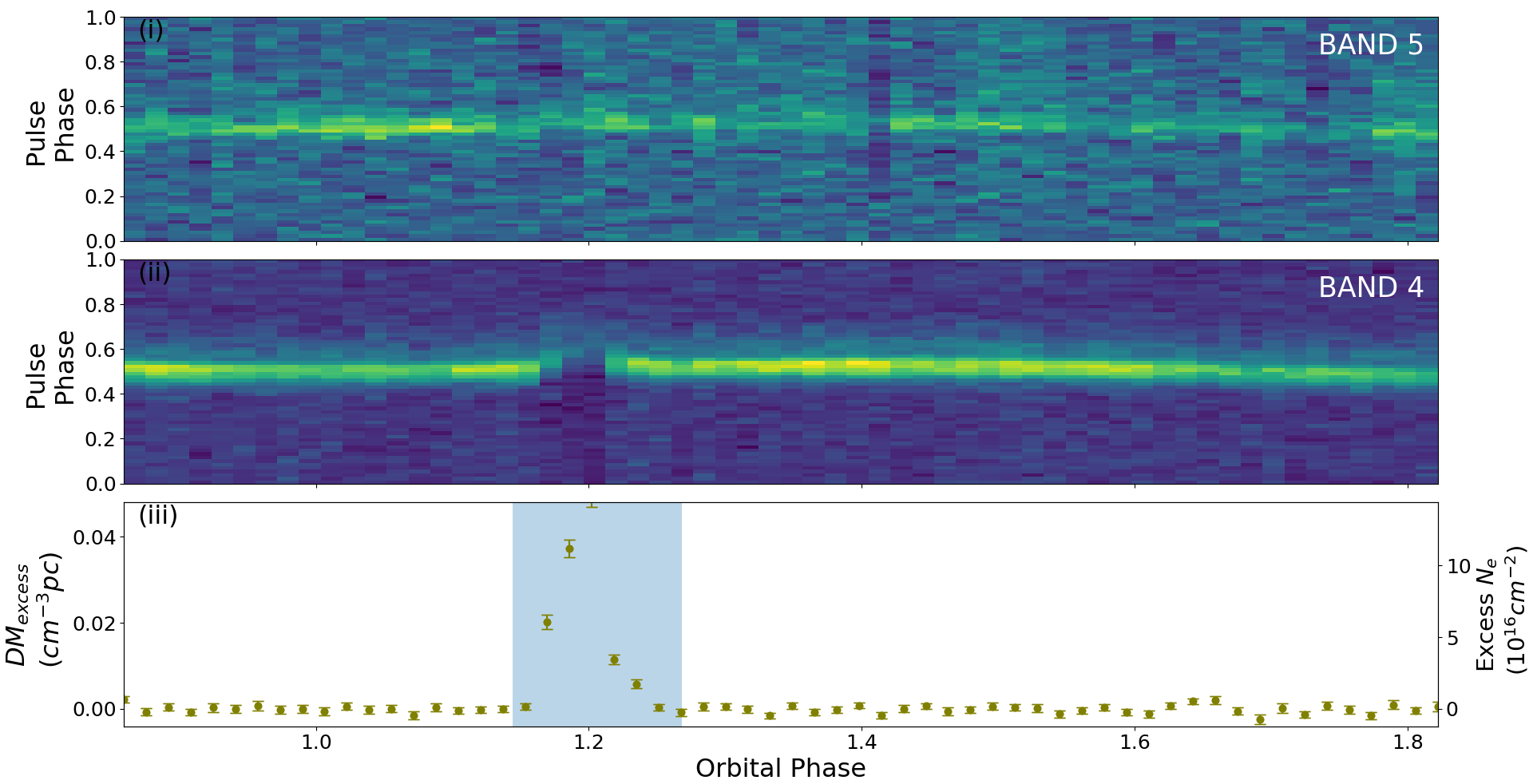}
         \textbf{(b)} Second eclipse for PSR J2051$-$0827
         
     \end{minipage}
     \hfill
     \begin{minipage}[b]{0.7\textwidth}
         \centering
         \includegraphics[width=\textwidth]{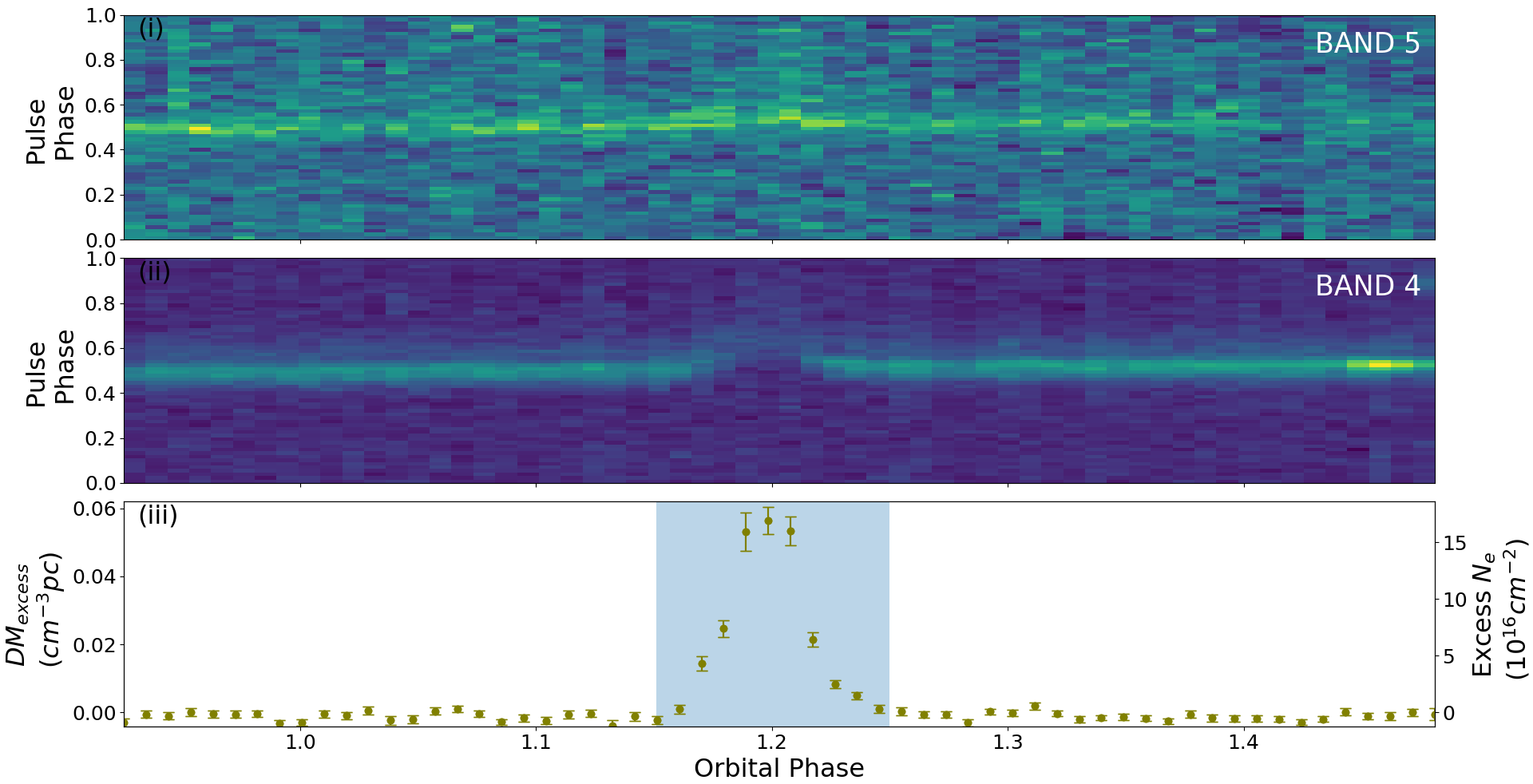}
         \textbf{(c)} Third eclipse for PSR J2051$-$0827
         \label{J2051_third_eclipse}
     \end{minipage}
    \caption{ Three consecutive eclipses plots for PSR J2051$-$0827. Three panels, from top to bottom depict variation of (i) the total intensity at band 5, (ii) the total intensity at band 4, (iii)  excess DM ($DM_{excess}$) along the line of sight at band 4. The eclipse phase is indicated by the shaded blue region.}
        \label{J2051_orbital_phase_resolved_plots}
\end{figure*}

\subsubsection{PSR J1555$-$2908}
\cite{J1555-2908_eclipse} found that this MSP eclipses at 820 MHz for 10$\%$ of the orbit, using Green Bank Telescope 200 MHz bandwidth GUPPI system. Therefore we observed this MSP in band 3, band 4 and band 5 of uGMRT, as the cutoff frequency would either be in band 5 or would be above that. With the uGMRT observations at band 3 and band 4, we detected the pulsar and observed eclipsing around the superior conjunction, as shown in Figure \ref{J1555_orbital_phase_resolved_plots}. Although the pulsar was detected in band 4, it was not detected in simultaneous band 5 observations. The theoretical 5-sigma sensitivity of the GMRT at 1.4 GHz in the direction of the pulsar is 13 µJy for 30 minutes of observation. The flux density reported by \cite{J1555-2908_eclipse} for PSR J1555--2908 at 1.5 GHz is between 0.15 and 0.24 mJy. Therefore, the pulsar should have been detected in band 5. The non-detection of PSR J1555--2908 in band 5 could be due to a temporal change in the flux densi
ty. We could not detect a precise value of the eclipse cutoff frequency for this system as we found complete eclipsing in band 3 and band 4 and conclude that it would be above band 4, that is $>$ 740 MHz. 

We also observed that flux density decreases near the eclipse boundary along with the rise in the DM near the eclipse boundary as can be noted from Figure \ref{J1555_orbital_phase_resolved_plots}, which was also noted for other BW MSPs. The corresponding value of $N_{e}$ is given in Table 1.

\subsubsection{PSR J2051$-$0827}
Previous eclipse studies have been carried out by \cite{PolzinJ2051} for PSR J2051$-$0827. They analysed the data, spanning frequencies from 110 MHz up to 4034 MHz using data from different radio telescopes (LOFAR, WSRT, Lovell, GMRT, Parkes). Variable eclipsing at 1 GHz was noted by \cite{PolzinJ2051} for PSR J2051$-$0827. In their study, 1.4 GHz radio emission is generally detected throughout the orbit, with only sporadic, short-duration dips in flux density and the corresponding TOA delays in some observations. However, no estimate of the eclipse cutoff frequency was mentioned.

We analyzed the band 4 and simultaneous band 5 data for this MSP to determine the eclipse cutoff frequency. Figure \ref{J2051_orbital_phase_resolved_plots} shows the detection plot for three consecutive eclipses of PSR J2051$-$0827. We obtained similar constraints on the eclipse cutoff frequency for this MSP in three consecutive eclipses observed. We noted that the eclipse cutoff frequency is below band 4 ($<$ 640 MHz, as the lower part of the band from 550--640 MHz, was hugely affected by the RFI) for the three consecutive eclipses covered in our observations.

\subsubsection{Mass loss rate estimation}
We estimated the mass loss rates of the companions for PSR J1810+1744 and PSR J1555--2908 in our sample, for which we report eclipses at band 4 of the uGMRT. The estimated mass loss rates are provided for both pulsars in Table 1. For PSR J1555--2908, the mass loss rate on 04 July 2020 is $6.5 \times 10^{-13}$ $\mathrm{M_\odot/year}$. However, we could not determine the mass loss rate on 29 June 2020 for this pulsar, as the complete eclipse was not covered. For PSR J1810+1744, we estimated the mass loss rates to be $9.0 \times 10^{-13}$ $\mathrm{M_\odot/year}$ and $6.5 \times 10^{-13}~ \mathrm{M}_\odot/year$ for the first and second eclipses observed on 26 June 2020, respectively. The estimated mass loss rates for both pulsars are not sufficient to evaporate the companion within the Hubble timescale, as is also reported for other spider MSPs. For instance, for PSR J2051$−$0827, the mass loss rate ($\dot{m}_{c}$) is $10^{-14} \mathrm{M_\odot}$/year \citep{stappers1996probing}
; for PSR J1959+2048 and PSR J1816+4510, $\dot{m}_{c}$ is $10^{-12}\mathrm{M_\odot}$/year and $2 \times 10^{-13} \mathrm{M_\odot}$/year \citep{polzin2020study}, respectively; and for PSR J1544+4937, $\dot{m}_{c}$ is $1.9 \times 10^{−14} \mathrm{M_\odot}$/year \citep{kumari2024}. The mass loss rate of the companion required for the complete ablation within the Hubble timescale needs to be greater than 4$\times 10^{−12} \mathrm{M_\odot}$/year for a initial companion mass of 0.05 $M_\odot$. Although estimated mass loss rate of around $10^{-12}\mathrm{M_\odot}$/year also exists for BW systems, \cite{PolzinJ1810} argued that this rate could be inconsistent, as mass loss from the binary system would increase the separation between the pulsar and its companion. Additionally, the spin-down of the pulsar would affect the $\dot{E}/a^{2}$ value, thereby reducing the likelihood of complete ablation. Therefore, BW MSP systems may not be the progenitors of isolated MSPs.

\subsection{Non detection of the low frequency eclipses}
Eclipse is not reported for PSR J0751$+$1807, PSR J1738$+$0333 and PSR J1807$-$2459A in previous studies.
We observed PSR J0751$+$1807 and PSR J1738$+$0333 with the uGMRT simultaneous at band 3 (300 MHz$-$500 MHz) and band 4 (550 MHz$-$750 MHz) during the eclipse phase (orbital phase $\sim$ 0.1$-$0.3). Possible white dwarf companions for these two systems predicted by \cite{J0751+1807_white_dwarf} and \cite{J1738+0333_white_dwarf} indicate that these may not belong to spider MSP category. MSPs in compact binaries with white dwarf companions do not show eclipses, as it is harder to expel material from the white dwarf companion. However, \cite{Khechinashvili_white_dwarf_eclipse_J1959_previous_stud} suggested an eclipsing mechanism for pulsars with white dwarf companions. In their model, the magnetic field of the white dwarf companion traps relativistic particles from the pulsar wind, giving rise to an extended magnetosphere around the companion star. The incoming radio waves from the pulsar undergo cyclotron damping as they pass through this magnetosphere, resulting in an eclipse.

We also observed PSR J1807$-$2459A at band 3, band 4, and band 5 of the uGMRT to investigate the presence of frequency-dependent eclipsing in this system. In the following section, we discuss the possible reason for non-detection of low frequency eclipses in individual pulsars in detail.

\subsubsection{PSR J0751+1807}
Eclipsing was not  reported from observations using Arecibo telescope for PSR J0751+1807 by \cite{J0751+1807_eclipse_previous} at 430 MHz (8 MHz bandwidth). Therefore, we observed this system simultaneously at band 3 and band 4 of the uGMRT to detect any potential eclipsing. Since eclipses are frequency-dependent and more pronounced at lower frequencies, sensitive observations with the uGMRT were expected to provide insights. However, we did not detect any eclipsing in this system at either band 3 or band 4.  The inclination angle for this system is $66^{+4}_{-7} $ degrees \citep{J0751_inclination}. At such an inclination angle, an eclipse should have been observable if there were sufficient ablated material around the companion. We conjecture that the non-detection of an eclipse could be due one or more of the following reasons. First, the frequency-dependent eclipsing might occur below band 3. Second, the companion may not be filling its Roche lobe. Third, the non-
detection could be attributed to the presence of the white dwarf companion \citep{J0751+1807_white_dwarf}.

\subsubsection{PSR J1738+0333}
An eclipse has not been reported for this MSP in previous study at around at 1 GHz \citep{J1738+0333_previous_study}. However, no low-frequency study is reported for this MSP, so we observed it at band 3 and band 4 of the uGMRT. However, we did not detect any eclipsing in this system at either band. The following reasons could lead to the non-detection of the eclipse in this system. First, it could be due to the low inclination angle of the system, which is $32^{+1}_{-1} $ degrees \citep{J1738+0333_white_dwarf}. Second, the reason could be frequency-dependent eclipsing and eclipse might be below band 3. Third, the companion may not be filling its Roche lobe. Fourth, the non-detection could be because the companion is a white dwarf.

\subsubsection{PSR J1807$-$2459A}
In previous studies, no eclipse was reported for this system by \cite{J1807-2459_scott_1.3_study} at 1.3 GHz, and also at 2 GHz by \cite{J1701-3006E_timing_J1807_nondetection_Eclipse}. We analyzed the simultaneous band 2 and band 3 data, the band 3 data using full array, as well as the simultaneous band 4 and band 5 data, as mentioned in Table \ref{tab:Table1}. We detected this pulsar in bands 3, 4, and 5 of the uGMRT. The pulsar signal is clearly visible in bands 4 and 5 during the eclipse phase. However, the detection significance in band 3 is low, which could be attributed to interstellar scattering caused by free electron density along the line of sight to the pulsar (DM $\sim$ 134 $\mathrm{pc ~cm^{-3}}$). We calculated the scattering timescale using the relation given by \cite{scattering_width}. We estimated it to be approximately 2 ms at band 3, which is close to the spin period of the pulsar, suggesting that scattering is dominant at band 3. Scattering could also be re
sponsible for the non-detection of the pulsar at band 2, where the scattering timescale is approximately 25 ms ($>>$ spin period of the pulsar).

\subsection{MSP not detected with the GMRT}
In this section, we will discuss the two MSPs in our sample that are not detected in either band 3 or band 4 of the uGMRT.

\subsubsection{PSR J1431$-$4715}
Signatures of eclipsing have been previously reported for PSR J1431$-$4715 at 1.4 GHz and 3 GHz by \cite{J1431_eclipse_study}. However, the authors noticed that the signal does not completely disappear at these frequencies. \cite{kumari_et_al_polarisation} also determined the eclipse cutoff frequency for this pulsar to be 1250 $\pm$ 80 MHz using the Parkes Telescope. Frequency-dependent eclipsing has been observed for this MSP in prior studies, noting that eclipses are more pronounced at lower frequencies \citep{J1431_eclipse_study}. 
We observed this MSP in band 3 and band 4 of uGMRT as shown in Table \ref{tab:Table1}. We could not detect this MSP in either band 3 or band 4. \cite{kumari_et_al_polarisation}, using the UWL receiver of the Parkes telescope, determined that the eclipse width, based on the rise and fall of dispersion measure (DM) around the superior conjunction ($\sim$ 0.25), extends from an orbital phase of 0.15 to 0.37 at 1.4 GHz. Consequently, in uGMRT band 3 and band 4, the eclipse should cover a larger fraction of the orbit. In our observations, we have covered the orbital phase from 0.2 to 0.3, and it is likely that we have sampled the orbit only during complete eclipse phase leading to non-detection at 400 MHz and 650 MHz.

\subsubsection{PSR J2215+5135}
Eclipses has previously been reported for this MSP by \cite{J2215+5132_discovery} at 350 MHz using the GBT. Detailed eclipse study using the inteferometeric observations is done by \cite{J2215-5135_eclipse} at 110--240 MHz and 30--80 MHz respectively using the LOFAR telescope. \cite{J2215-5135_eclipse} reported frequency dependent eclipsing for PSR J2215+5135 ($\propto \nu^{-0.4} $, where $\nu$ is the frequency). According to \cite{J2215-5135_eclipse}, the full width at half maximum (FWHM) of the eclipse is given by, $\mathrm{FWHM = (0.476 \pm 0.026)(\nu/150)^{(-0.42\pm0.11)}}$. This suggests that the eclipse is expected to last for around 31$\pm$8\% of the orbital period at 400 MHz, corresponding to the FWHM. The orbital phase we observed covers 0.04 to 0.4, which is about 40\% of the orbit. Therefore, non-detection could be hugely affected by eclipsing, for PSR J2215+5135.

\subsection{Eclipse width as the function of frequency}
We investigated how the eclipse width varies as a function of frequency by dividing the 400 MHz wide bandwidth uGMRT band 4 data (with a usable bandwidth of approximately 300 MHz) into smaller chunks for PSR J1810+1744 and PSR J1555--2908. For PSR J2051--0827, the sensitivity was not sufficient to split the 400 MHz bandwidth into chunks and study the frequency evolution. The procedure followed to determine the eclipse width is detailed in Section \ref{subsec:eclipse parameter estimation}. For PSR J1810+1744, the analysis was conducted for the consecutive main eclipses observed in our dataset (around 0.25 orbital phase) and the eclipse occurring at a non-standard orbital phase (around 0.5). Our study did not reveal any significant frequency evolution of the eclipse width, either for the main eclipse or for the eclipse at the non-standard orbital phase, for PSR J1810+1744. Similarly, we did not observe any frequency evolution of the eclipse width for PSR J1555--2908 on 4th July
 2020, as complete eclipse phase was covered during this epoch. The eclipse width was calculated to be $\sim$ 16 minutes for PSR J1810+1744 and $\sim$ 36 minutes for PSR J1555--2908. The eclipse width may vary very slowly within the observed band, with changes smaller than the length of a sub-integration in time. The length of the sub-integration is $\sim$ 2.4 minutes for PSR J1555--2908 and $\sim$ 1.7 minutes for PSR J1810+1744. Assuming that the scaling of the eclipse width ($E_{w}$) with frequency is $E_{w}=K\nu^{-\alpha}$, and given that the eclipse width could not change more than the length of a sub-integration, we constrained the value of $\alpha$ to be less than 0.22 for PSR J1810+1744, and 0.20 for PSR J1555--2908.

Previous studies conducted by \cite{PolzinJ1810} reported the value of $\alpha$ to be 0.41 $\pm$ 0.03 for PSR J1810+1744. The difference between our value and the value estimated by \cite{PolzinJ1810} could be attributed to the different methods employed to measure the eclipse width. In their study, \cite{PolzinJ1810} measured the eclipse width as the width of the eclipse at half of the out-of-eclipse flux density. Additionally, \cite{Fruchter1988a} and \cite{J2215-5135_eclipse} found $\alpha$ to be 0.41 for PSR J1959+2048 and PSR J2251+5135, respectively. However, for PSR J1720--0533, \cite{J1720-0533_Wang_depolarisation}, using the FAST telescope (1.05--1.45 MHz), concluded that the eclipse duration is not monotonically decreasing with increasing frequency due to the effect of scintillation, which makes it difficult to confirm where the eclipse begins or ends. \cite{J1720-0533_Wang_depolarisation} reported the value of $\alpha$ for J1720--0533 to be 0.14 $\pm$ 0.07.

\cite{Thompson1992} proposed several eclipse mechanisms to explain the frequency-dependent eclipsing observed in spider MSPs. Each mechanism results in a different evolution of eclipse width with frequency (i.e., different values of $\alpha$). Therefore, by determining the value of $\alpha$, the eclipse mechanism in PSR J1810$+$1744 can be probed.
Additionally, we did not observe any temporal variation in the eclipse width for the two consecutive main eclipses covered in our observations.

\section{Probing the possible cause for eclipsing}
\label{sec:possible cause of eclipsing}
The $\dot{E}/a^{2}$ values (sourced from the ATNF Pulsar Catalogue), along with the inclination angle, RLFF, and mass function (taken from the parameter file available at the ATNF Pulsar Catalogue), are provided for compact MSP binaries in Table \ref{tab:Table3}. These binaries have orbital periods shorter than one day and median companion masses below 0.8 $M_{\odot}$.
Although the majority of BW MSPs exhibit eclipses, approximately 50\% of them have no previously reported eclipses (as listed in Table \ref{tab:Table3}). However, it must be noted that low-frequency studies exist for only $\sim$ 36\% of the BW MSPs where no eclipses have been reported. Therefore, the possibility of eclipses in the remaining $\sim$ 67\% of BW MSPs cannot be ruled out. The absence of eclipses in spider MSPs can be attributed to factors such as low orbital inclination angles, low Roche lobe filling factors and low $\dot{E}/a^{2}$ values. In the following sections, we will discuss the dependence of eclipses on each of these parameters.

\subsection{Role of Inclination angle and Mass function}
For the radio beam of the pulsar to pass through the ablated material, a high orbital inclination angle is necessary. Thus, eclipses in spider MSPs are generally expected in systems with higher orbital inclinations. The inclination angle can be determined either through timing analysis, by estimating post-Keplerian parameters, or through optical light curve modeling. An estimation of inclination angle exists for approximately $\sim$ 28 \% of spider MSPs (can be noted from Table \ref{tab:Table3}). Relatively, lower inclination angles, such as those determined for PSR J0023+0923, J0636+5129, J0952--0607, and J1301+0833 can explain the absence of eclipses in these systems. On the contrary there are instances where systems with low orbital inclination angles exhibit eclipses. Examples include PSR J0251+2606, J1124--3653, J1544+4937, J1641+8049, J1723--2837, and J1810+1744, where eclipses are present despite the low inclination angles. 

As detailed in the previous paragraph, inclination angle estimates are available for only a limited number of spider MSPs. Therefore, another parameter, the mass function, serves as a useful tool for indirectly studying the correlation between the inclination angle and the presence of eclipses.
The mass function of a binary system is proportional to the $\sin^3(i)$, where i is the orbital inclination angle. Consequently, the mass function is expected to be higher in eclipsing spider binaries, as these systems typically have higher orbital inclination angles. This assumption holds true when the companion mass and the pulsar mass are approximately the same across the BW and RB MSP populations. \cite{Freire_2005_J0024} studied the correlation of mass function with the eclipsing in spider MSPs, and found that the eclipsing binaries tend to have a higher mass function.  
Similarly, \cite{massfunction_eclipse} used one-dimensional Kolmogorov–Smirnov (KS) tests \citep{KS_test} and concluded that the probability of eclipsing and non-eclipsing objects originating from the same parent distribution is very low. They noted that eclipses are generally observed in systems with a higher mass function \citep[as shown in Table 2 of][]{massfunction_eclipse}. We also conducted a KS test for eclipsing and non-eclipsing BW MSPs, using a sample of 54 BW MSPs consisting of 24 non-eclipsing and 30 eclipsing BW MSPs. We obtained a KS statistic of 0.60 with a p-value of 0.00005, indicating that eclipsing and non-eclipsing BWs are likely drawn from distinct distributions. This suggests that eclipsing BW MSPs tend to have higher mass functions compared to the non-eclipsing BW MSP population, as reported by \cite{Freire_2005_J0024} and \cite{massfunction_eclipse}.

\subsection{Role of Roche lobe filling factor}
Mass loss from the binary companion can occur from three primary reasons. First, the energetic pulsar wind can directly ablate mass from the companion. Second, evaporative mass loss \citep{companion_winds_nature,companion_wind_nature_2} can result from heating effects caused by the energetic pulsar wind and $\gamma$-ray emission from the intra-binary shock region \citep[observed for few spider MSPs,][]{Intrabinary_shock_J1311-3430,Intrabinary_shockJ2241-5236}. The heating of the companion may cause it to expand, potentially filling its Roche lobe and leading to Roche lobe overflow, which constitutes the third cause of mass loss.  
The Roche lobe filling factor quantifies how much of the Roche lobe a star occupies. It is defined as the ratio of the radius of the star to its Roche lobe. A filling factor close to 1 indicates that the outer layers of the star are loosely bound, allowing for mass transfer or the easy ablation of material from the companion \citep{Mata_optical_J1544}. For spider MSPs, the Roche lobe filling factor can be determined through optical light curve modeling, as the companion stars are often visible in optical wavelengths. Roche lobe filling factor data are available for only a small fraction of spider MSPs ($\sim$ 23 \%) primarily for eclipsing systems as is tabulated in Table \ref{tab:Table3}. There are few systems with relatively higher value of RLFF, for which eclipses are not reported. For example eclipses are not seen for PSR J0636+5129 and PSR J0952--0607 with RLFF of $>$ 0.95 and 0.87 respectively. Any possible trend of RLFF with eclipsing can only be investigated w
hen RLFF becomes available for a larger sample of spider MSPs in the future.

\subsection{Role of $\dot{E}/a^{2}$}
$\dot{E}/a^{2}$ is measure of how much the pulsar wind and energetic radiation would be effective at the position of the companion. $\dot{E}/a^{2}$ value was available for a sample of 39 eclipsing spider MSPs and for a sample of 22 non-eclipsing spider MSPs. Out of the 22 non-eclipsing spider MSPs, low-frequency studies have been reported for only $\sim$ 8 systems. For the rest of the pulsars in the sample, no low-frequency studies (around 300 MHz) have been reported. Therefore, it would not be correct to completely rule out the presence of an eclipse in these systems. 
We conducted a KS test to determine whether the $\dot{E}/a^{2}$ values are of the same order for both the spider MSP population that shows eclipses and those that do not. Our sample consisted of 8 non-eclipsing (where low frequency observations are reported) and 39 eclipsing spider MSPs. The test yielded a KS statistic of 0.41 and a p-value of 0.16. Since the p-value is higher than typical significance thresholds ($\sim$ 0.05), the test does not provide strong evidence that the two distributions are different.
This indicates that the other factors such as the orbital inclination angle and Roche lobe filling factors could be playing a significant role for these systems. From Table \ref{tab:Table3}, it can be observed that there are a few BW MSPs where the $\dot{E}/a^{2}$, inclination angle, and Roche lobe filling factor are sufficient to suggest the presence of an eclipse, yet no eclipse is observed. Examples include PSR J0610--2100 and PSR J0952--0607.

\section{Summary}
\label{sec:summary}
We present a low-frequency eclipse study of a sample of 10 millisecond pulsars (MSPs) using the GMRT. Our study reports the first detection of frequency-dependent eclipsing in two spider MSPs, PSR J2214+3000 and PSR J2234+0944, observed in one epoch but not in subsequent epochs, indicating potential variability in the cutoff frequency over time.
We aimed to determine the eclipse cutoff frequency for three spider MSPs in our sample: PSR J1555--2908, PSR J1810+1744, and PSR J2051-0827. We found that the cutoff frequency for PSR J1555-2908 and PSR J1810+1744 lies above GMRT band 4, while for PSR J2051--0827, it lies below band 4. Notably, we detected an eclipse at an orbital phase of 0.55 for PSR J1810+1744, the first such detection at a non-standard orbital phase for this pulsar. Mini-eclipses at non-standard orbital phases have been observed in other pulsars, but the eclipse at 0.55 lasted longer than the main eclipse at 0.25, which is unprecedented for spider pulsars.
We estimated the mass loss rates for PSR J1810+1744 and PSR J1555--2908, finding them insufficient for complete companion ablation within reasonable timescales. We did not detect eclipses in PSR J0751+1807, PSR J1738+0333, and PSR J1807--2459A. The absence of eclipses in PSR J0751+1807 could be due to the companion being a white dwarf, not filling its Roche lobe, or the eclipse occurring below GMRT band 3. For PSR J1738+0333, the non-detection could be due to a white dwarf companion, low orbital inclination angle, non-filling Roche lobe, or eclipses below GMRT band 3. In PSR J1807--2459A, the reasons could be a low inclination angle, eclipses below band 3, or the companion not filling its Roche lobe.
We did not detect PSR J1431--4715 and PSR J2215+5135. The non-detection of PSR J1431--4715 could be due to a complete eclipse covered during the observation, and for PSR J2215+5135, it could be due to the pulsar's variable flux density falling below GMRT's detection limit.
We also examined the frequency evolution of eclipse width for PSR J1555--2908 and PSR J1810+1744, finding no evolution with frequency. Additionally, using the KS statistic, we found that the $\dot{E}/a^{2}$ values are of the same order for both the spider MSP populations that show eclipses and those that do not. This suggests that higher $\dot{E}/a^{2}$ values do not necessarily guarantee eclipses, and that other factors, such as the Roche lobe filling factor (RLFF), orbital inclination angle, and the nature of the companion, also play a role. However, additional estimates for RLFF and inclination angle values for a larger sample of spider MSPs would help clarify which specific combinations of these factors are necessary for eclipsing to occur. We also concluded that the eclipsing BW MSPs tend to have higher mass function value compared to the non-eclipsing MSPs.

\begin{table*}[hbt!]
\begin{center}
\caption{Table listing the $\dot{E}/a^{2}$, orbital inclination angle and Roche lobe filling factor (RLFF) values for the compact MSPs binaries.}
\label{tab:Table3}
\vspace{0.3cm}
\begin{longtable}[c]{ccccccc}
\hline
Pulsar name & Eclipse& $\dot{E}/a^{2}$ & Inclination& RLFF & Mass & References\\
& & ($\times 10^{35} erg/s $\(R_\odot\)$^{2}$) & (degrees) & & function&\\
\hline
J0023+0923 (BW) & no &0.09&  $42^{+4}_{-3}$ & $0.36^{+0.18}_{-0.13}$ &0.000002 &\cite{bak2020timing}\\
& & & & & & \cite{Mata_optical_J1544} \\
J0024--7204I (BW) & no &  & & &0.000001& \cite{Freire200347tuc}\\
\bf{J0024--7204J (BW)} & yes&& & &0.000005& \cite{J0024_previous_study}\\
\bf{J0024--7204O (BW)} & yes &0.39  & & &0.000005& \cite{J0024_previous_study}\\
J0024--7204P (BW)& no &2.96& & &0.000003& \cite{47_Tuc_eclipsing_pulsars}\\
\bf{J0024--7204R (BW)} & yes &2.22 & & &0.000009& \cite{J0024_previous_study}\\
J0024--7204U (WD)& no &0.05& & & 0.000853& \cite{47_Tuc_eclipsing_pulsars}\\
\bf{J0024--7204V (RB)} & yes & & & &0.009744&\cite{47_Tuc_eclipsing_pulsars}\\
\bf{J0024--7204W (RB)} &yes& & & &0.000876& \cite{J0024_previous_study}\\
J0024--7204Y (WD)& no & & & &0.001178 & \cite{J0024_previous_study}\\
\bf{J0024--7204ac (BW)} & yes& & & &0.000000&\cite{J1701-3006G_timing_47_Tuc_ac_ad} \\
\bf{J0024--7204ad (RB)})& yes & & & &0.003297& \cite{J1701-3006G_timing_47_Tuc_ac_ad}\\
\bf{J0251+2606 (BW)} & yes &0.06& $32^{+2}_{-1}$& $0.60^{+0.07}_{-0.06}$ &0.000007& \cite{Mata_optical_J1544}\\
J0312--0921 (BW)& - & 0.14 & & & 0.000000 & \cite{J0312-0921_black_widow}\\
J0610--2100 (BW)& no & 0.01 &54  $<$ i $<$ 89 &  0.72 $< f_{RL} <$ 0.94&0.000005& \cite{J0610-2100_non_Eclipse}\\
J0636+5129 (BW)& no & 0.09 & 24 $\pm$ 1 & $>$ 0.95 & 0.000000 & \cite{J0636+5129_no_eclipse}\\
& & & & & & \cite{Mata_optical_J1544} \\
J0751+1807 (WD) & no &0.03 & $66^{+4}_{-7}$ & & 0.000967& \cite{J0751+1807_white_dwarf} \\
& & & & & & \cite{J0751_inclination} \\
J0952--0607 (BW) & no & 0.16 & $56^{+5}_{-4}$ & $0.87^{+0.02}_{-0.03}$ & 0.000004 & \cite{J0952-0607_no_Eclipse}\\
& & & & & & \cite{Mata_optical_J1544} \\
J1012+5307 (WD)& no &0.003& 50 $\pm$ 2& &0.000578& \cite{J1012+5307_WD}\\
\bf{J1023+0038 (RB)}& yes & 0.19& & &0.001107& \cite{J1023+0038_eclipse}\\
\bf{J1036--4353 (RB)}& yes & & & &0.004707& \cite{J1036-4353_eclipse_J1526-2744_wd}\\
J1036--8317 & no& & & &0.001239& \cite{J1946-5403_J1036-8317_no_eclipse}\\
\bf{J1048+2339 (RB)} & yes & 0.02 & $<$ 76 &0.83 $\pm$ 0.03& 0.010000 & \cite{J1048+2339_inclination_roche}\\
\bf{J1124--3653 (BW)} & yes & 0.05&  44 & 0.84 & 0.000011 & \cite{J1124-3653_roche_inclination}\\
\bf{J1221--0633 (BW)} & yes & 0.04 & & & 0.000001 &  \cite{J1221-0633_eclipse}\\
\bf{J1227--4853 (RB)} & yes & 0.18 &77 & 0.83 & 0.003870 & \cite{J1227-4853_roche_lobe}\\
J1301+0833 (BW) & - & 0.15& 52 &0.57 &0.000007& \cite{J1301+0833_inclination_Roche_lobe}\\
\bf{J1302--3258 (RB)} & yes & 0.002 & & & 0.001394 & \cite{J1302-3258_eclipse}\\
\bf{J1311--3430 (BW)} & yes & 0.8 & 60 & 0.99 & 0.000000 & \cite{J1311-3430_inclination}\\
\bf{J1317--0157 (BW)} & yes & 0.09 & & & 0.000003 & \cite{J1221-0633_eclipse}\\
J1326--4728B & - & & & & 0.000001 & \cite{J1326-4728B_timing}\\
J1342+2822A (BW) & no& & & &0.000001& \cite{J1342+2822A_no_Eclipse}\\
\bf{J1431--4715 (BW)} & yes & 0.07 & 53--83 & 0.64--0.76 & 0.000883 & \cite{J1431_inclination_roche}\\
J1446--4701 (BW) & no &0.08& & &0.000004& \cite{J1446-4701_no_eclipse}\\
\bf{J1513--2550 (BW)} & yes & 0.38 & 70 & 1 & 0.000002 & \cite{Sanpa-arsa_phd_thesis}\\
 &  & & & & & \cite{J1446-4701_no_eclipse}\\
 &  & & & & &\cite{J1513-2550_inclination_roche} \\
\bf{J1518+0204C (BW)} & yes & 0.73& & &0.000027& \cite{J1518+0204C_eclipse}\\
J1518+0204G (BW) & no & 0.18 & & & 0.000004 & \cite{J1518+0204G_no_eclipse}\\
J1526--2744 (WD) & no & 0.03 & & & 0.000294 & \cite{J1036-4353_eclipse_J1526-2744_wd}\\
\bf{J1544+4937 (BW)} & yes & 0.07 &$47^{+7}_{-4}$ & $>$0.96 & 0.000003 &  \cite{Mata_optical_J1544}\\
\bf{J1555--2908 (BW)} & yes &  0.90 & $>$ 75 & 0.98 & 0.000068 & \cite{J1555_inclination} \\
J1602--1009 & - & 0.04 & & & 0.000003 & \cite{J1602-1009_}\\
\bf{J1622--0315 (RB)} & yes&0.03&$<$ 83.4 & &0.000433& \cite{Sanpa-arsa_phd_thesis}\\
&  & & & & & \cite{gamma_ray_eclipse}\\
\hline
\end{longtable}
\end{center}
\end{table*}

\begin{table*}[hbt!]
\begin{center}
\vspace{0.3cm}
\begin{longtable}[c]{ccccccc}
\hline
Pulsar name  & Eclipse &$\dot{E}/a^{2}$ & Inclination& RLFF & Mass function& References\\
& & ($\times 10^{35} erg/s $\(R_\odot\)$^{2}$) & (degrees) & & &\\
\hline
J1627+3219 (BW)& - &0.09& & &0.000006& \cite{J1627+3219_BW}\\
\bf{J1628--3205 (RB)} & yes &0.04&$>$ 55 & &0.001681& \cite{J1628-3205_eclipse}\\
J1630+3550 (BW) & no &0.05& & &0.000001& \cite{J1630+3550_no_eclipse}\\
J1641+3627D (WD) &no & & & &0.002421& \cite{J1641+3627D_whitedwarf}\\
\bf{J1641+3627E (BW)} & yes &0.35 & & & 0.000004 & \cite{J1641+3627E_eclipse}\\
\bf{J1641+8049 (BW)} & yes & 0.00004& 57 $\pm$ 2 &  $>$ 0.95 & 0.000034 & \cite{Mata_optical_J1544}\\
J1653--0158 (BW)& - &0.26& $72.3^{+5}_{-4.9}$ & $0.88^{+0.03}_{-0.03}$ & 0.000000 & \cite{J1653-0158_gamma_ray_obs}\\
\bf{J1701--3006B (RB)} &yes & & & &0.000830& \cite{J1701-3006B_eclipse}\\
J1701--3006C (BW)& no & & & &0.000167& \cite{J1701--3006C_timing}\\
\bf{J1701--3006E (BW)} & yes &1.78 & & &0.000015& \cite{J1701-3006E_timing_J1807_nondetection_Eclipse}\\
J1701--3006F (BW) & no& 2.56 & & & 0.000005 & \cite{J1701-3006E_timing_J1807_nondetection_Eclipse}\\
J1701--3006G (WD) & no & & & & 0.000427 & \cite{J1701-3006G_timing_47_Tuc_ac_ad}\\
\bf{J1717+4308A (RB)} &yes& 0.26& & &0.001687& \cite{J1717+4308A_eclipse}\\
J1719--1438 (WD) & no & 0.01& & &0.000000& \cite{J1719-1438_msp}\\
J1720--0533 (BW) & yes &0.05& & &0.000013& \cite{J1720-0533_Wang_depolarisation} \\
\bf{J1723--2837 (RB)} & yes& 0.03 &30--41 & & 0.005221& \cite{Crawford_J1723-2837_eclipse_study}\\
J1727--2951 (BW) & no &0.0007& & &0.000001& \cite{J1727-2951_no_Eclipse}\\
\bf{J1731--1847 (BW)} & yes& 0.15& & &0.000019& \cite{J1731-1847_eclipse}\\
\bf{J1737-0314D (RB)} & yes && & && \cite{J1737-0314D_eclipse}\\
\bf{J1737-0314E (RB)} & yes&& & && \cite{J1737-0314D_eclipse}\\
J1737--0314A (BW)& no &1.50& & & 0.000002 & \cite{J1737-0314D_eclipse}\\
J1738+0333 (WD) & no & 0.007 &32.6 & & 0.000346 & \cite{J1738+0333_white_dwarf}\\
J1745+1017 (BW)& no &0.003& & &0.000001& \cite{J1745+1017_no_Eclipse}\\
\bf{J1745--23 (BW)} &yes& & & &0.000010& \cite{J1727-2951_no_Eclipse}\\
\bf{J1748--2021D (RB)}& yes &0.02 & & &0.000822& \cite{J1748-2021D_eclipse}\\
\bf{J1748--2446A (RB)}& yes & & & & 0.000322 & \cite{J1748-2446A_eclipse}\\
J1748--2446M (WD) & no && & &0.001158& \cite{J1748-2446M_N_WD}\\
J1748--2446N (WD)& no && & &0.030660& \cite{J1748-2446M_N_WD}\\
\bf{J1748--2446O (BW)} & yes && & &0.000022& \cite{J1748-2446O_P_eclipse}\\
\bf{J1748--2446P (RB)} & yes && & &0.016807& \cite{J1748-2446O_P_eclipse}\\
J1748--2446V (WD)&no& & & &0.000772&\cite{J1748-2446M_N_WD}\\
J1748--2446ab & -& & & &0.000322& -\\
J1748--2446ae (BW)&-& & & &0.000002& \cite{J1748-2446M_N_WD}\\
J1748--2446ai & -&& & &0.032863&-\\
J1748--2446am (WD) &-& & & &0.001383& \cite{J1748-2446am_WD}\\
J1753--2819 & -& & & &0.000066& \cite{J1753-2819}\\
J1757--5322 (WD)& no &0.001 & & &0.047464& \cite{J1757-5322_WD}\\
\bf{J1803--6707 (RB)} & yes&0.01& & &0.008882& \cite{J1036-4353_eclipse_J1526-2744_wd}\\
J1805+0615 (BW)& -&0.16 & & &0.000006& \cite{J0251+2606_eclipse}\\
J1807--2459A (BW) &no & & & &0.000000& \cite{J1701-3006E_timing_J1807_nondetection_Eclipse}\\
\bf{J1810+1744 (BW)} & yes &0.21& 48 $\pm$ 7 & 0.80 $\pm$ 0.3 &0.000041& \cite{J0023+0923_rochelobe}\\
\bf{J1816+4510 (RB)} & yes &0.08& & &0.001740& \cite{J0636+5129_no_eclipse}\\
J1824--2452G (BW)& - & & & &0.000000& \cite{J1824-2452G_black_widow}\\
\bf{J1824--2452H (RB)}& yes & & & & 0.002113 &\cite{J1824-2452G_black_widow}\\
\bf{J1824--2452I (RB)}&yes& & & &0.002286& \cite{J1824-2452I_eclipse}\\
J1824--2452J & -& & & &0.000002& - \\
J1824--2452L & -& & & &0.000004& -\\
J1824--2452M (BW)& no &0.12& & &0.000001& \cite{J1824-2452M_N_BW}\\
\hline

\end{longtable}
\end{center}
\end{table*}

\begin{table*}[hbt!]
\begin{center}
\vspace{0.3cm}
\begin{longtable}[c]{ccccccc}
\hline
Pulsar name  & eclipse&$\dot{E}/a^{2}$ & Inclination& RLFF & Mass function& References\\
& &($\times 10^{35} erg/s $\(R_\odot\)$^{2}$) & (degrees) & & &\\
\hline
J1824--2452N (BW)& no &0.62 & & &0.000003& \cite{J1824-2452M_N_BW}\\
J1833--3840 (BW) & - & 0.05& & &0.000000& \cite{gamma_ray_eclipse}\\
J1836--2354A (BW) & no &0.008& & &0.000003& \cite{J1836-2354A_no_eclipse}\\
J1850+0244 ($BW^{*}$)& no &0.04 & & &0.000184& \cite{J1850+0244_no_eclipse}\\
J1905+0154A & -& & & &0.000332& \cite{J1905+0154A_no_eclipse_sparse_Sampling}\\
J1906+0055 (WD) &-&0.004 & & &0.000705 & \cite{J1906+0055_wd}\\
\bf{J1908+2105 (RB)} & yes & 0.17 & & & 0.000080 & \cite{J0251+2606_eclipse}\\
J1910--5959A (WD)&-& & & & 0.002688 & \cite{J1910-5959A_WD}\\
J1911+0102A (BW)& no && & &0.000003& \cite{J1911+0102A_no_Eclipse}\\
J1928+1245 (BW)& no &0.14& & &0.000000& \cite{J1928+1245_no_eclipse}\\
J1946--5403 (BW)& no &0.05& & &0.000005& \cite{J1946-5403_J1036-8317_no_eclipse}\\
J1953+1844 & no &0.34& & &0.000000& \cite{J1953+1844_no_eclipse}\\
J1953+1846A (BW) & yes && & &0.000016& \cite{J1953+1846A_eclipse}\\
\bf{J1957+2516 (RB)} &yes &0.04& & &0.000431& \cite{J1906+0055_wd}\\
\bf{J1959+2048 (BW)} &yes& 0.24& & &0.000005& \cite{Khechinashvili_white_dwarf_eclipse_J1959_previous_stud}\\
J2006+0148 (WD)& -&0.009& & &0.001514& \cite{J0251+2606_eclipse}\\
\bf{J2017--1614 (BW)}& yes &0.07 & & &0.000009& \cite{Sanpa-arsa_phd_thesis}\\
\bf{J2039--5617 (RB)} & yes &0.08 & 61 $<$ i $<$ 78 & &0.002159& \cite{J2039-5617_eclipse}\\
J2047+1053 (BW)& -&0.06& & &0.000023& \cite{ray2012radio}\\
\bf{J2051--0827 (BW)} & yes &0.05& & &0.000010& \cite{stappers1996probing}\\
\bf{J2052+1219 (BW)} & yes&0.25& & &0.000019& \cite{J2052+1219_inclination}\\
\bf{J2055+1545 (RB)} & yes &0.12& & &0.005747& \cite{J1630+3550_no_eclipse}\\
\bf{J2055+3829 (BW)}& yes&0.02& & &0.000006& \cite{massfunction_eclipse}\\
\bf{J2115+5448 (BW)} &yes&1.04 & & &0.000005& \cite{Sanpa-arsa_phd_thesis}\\
\bf{J2129--0429(RB)} & yes& & 80.5$\pm$7.0 &0.95 $\pm$ 0.01 &0.016848& \cite{J2129-0429_inclination}\\
\bf{J2140--2310A (RB)} &yes& & & &0.000460& \cite{J2140-2310A_eclipse}\\
J2214+3000 (BW)& &0.02 & & &0.000001& \cite{bak2020timing}\\
\bf{J2215+5135 (RB)} & yes &0.30 & 66 $\pm$ 16 & 0.99 $\pm$ 0.03 & 0.003702 & \cite{J0023+0923_rochelobe}\\
\bf{J2234+0944 (BW)} & &0.02& & &0.000002& \cite{bak2020timing}\\
J2241--5236 (BW) & no& 0.14& & &0.000001& \cite{J2241-5236_no_eclipse}\\
\bf{J2256--1024 (BW)} & yes&0.12& 68 $\pm$ 11& 0.40 $\pm$ 0.20&0.000014&\cite{J0023+0923_rochelobe}\\
J2322--2650 (BW) & no &0.001& & &0.000000& \cite{J2322-2650_no_eclipse}\\
\bf{J2339--0533 (RB)} & yes&0.07& & &0.006589& \cite{J2339-0533_eclipse}\\
\hline

\end{longtable}
\end{center}
The pulsars written in the bold are those for which eclipse has been reported. The $\dot{E}/a^{2}$ and the mass function values are taken from the ATNF pulsar catalogue. The updated catalog of spider pulsar parameters is available at  \href {https://sangitakumari123.github.io/}{https://sangitakumari123.github.io/}. \\

\vspace{1cm}
\end{table*}

\section*{Acknowledgments}
We acknowledge support of the Department of Atomic Energy, Government of India, under project no.12-R\&D-TFR-5.02-0700. The GMRT is run by the National Centre for Radio Astrophysics of the Tata Institute of Fundamental Research, India. We acknowledge the support of GMRT telescope operators for observations. 
\clearpage\newpage
\bibliography{citation.bib}{}
\bibliographystyle{aasjournal}

\end{document}